\documentclass[aps,prl,twocolumn,preprintnumbers,groupedaddress]{revtex4}

\usepackage{amssymb}
\usepackage{epsfig} 
\usepackage{amsmath,color} 
\usepackage{graphicx} 
\usepackage{ulem} 
\usepackage{multirow}
\usepackage{tabularx}

\usepackage{wrapfig}

\usepackage{tikz}
\usetikzlibrary{positioning,arrows}
\usetikzlibrary{decorations.pathmorphing}
\usetikzlibrary{decorations.markings}

\newcommand{\beq}{\begin{equation}}
\newcommand{\eeq}{\end{equation}}
\newcommand{\bea}{\begin{eqnarray}}
\newcommand{\eea}{\end{eqnarray}}
\newcommand{\ba}{\begin{array}}
\newcommand{\ea}{\end{array}}
\newcommand{\bi}{\begin{itemize}}
\newcommand{\ei}{\end{itemize}}
\newcommand{\bn}{\begin{enumerate}}
\newcommand{\en}{\end{enumerate}}
\newcommand{\bc}{\begin{center}}
\newcommand{\ec}{\end{center}}
\renewcommand{\l}{\left}
\renewcommand{\r}{\right}

\newcommand{\eq}[1]{Eq.~(\ref{#1})}
\newcommand{\eqs}[2]{Eqs.~(\ref{#1}) and (\ref{#2})}
\newcommand{\eqss}[3]{Eqs.~(\ref{#1}), (\ref{#2}) and (\ref{#3})}

\newcommand{\GeV}{\mathinner{\mathrm{GeV}}}
\newcommand{\TeV}{\mathinner{\mathrm{TeV}}}

\begin{document}

\preprint{FTUV-16-05-07}
\preprint{IFIC-16-25}

\title{Gravitational waves from first order phase transitions as a probe of an early matter domination era and its inverse problem}


\author{Gabriela Barenboim}
\email[]{Gabriela.Barenboim@uv.es}
\author{Wan-Il Park}
\email[]{Wanil.Park@uv.es}
\affiliation{
Departament de F\'isica Te\`orica and IFIC, Universitat de Val\`encia-CSIC, E-46100, Burjassot, Spain}


\date{\today}

\begin{abstract}
We investigate the gravitational wave background from a first order phase transition in a matter-dominated universe, and show that it has a unique feature from which important information about the properties of the phase transition and thermal history of the universe can be easily extracted.
Also, we discuss the inverse problem of such a gravitational wave background in view of the degeneracy among macroscopic parameters governing the signal.
\end{abstract}


\maketitle


\section{Introduction}

In addition to cosmic microwave background (CMB) which provides a large mount of information about our universe \cite{Ade:2015xua}, gravitational waves produced by cosmological events can be very useful tools to probe the early history of the universe well before the epoch of Big Bang Nucleosynthesis.
Their interactions to the particle content of the universe are negligible and hence they are expected to leave clear imprints of the universe at the time of their generation.
In particular, \textit{inflation} \cite{Guth:1979bh,Kazanas:1980tx,Starobinsky:1980te,Guth:1980zm,Sato:1980yn,Sato:1981ds,Linde:1981mu,Albrecht:1982wi} which is believed to be the key ingredient of the very early history of our universe is expected to accompany a scale invariant gravitational wave background (GWB) which may be within the reach of direct detection experiments, for example BBO \cite{BBO} or DECIGO \cite{Kawamura:2011zz}.
Actually, there are various scenarios for stochastic GWBs: for example, inflation \cite{Starobinsky:1979ty}, preheating \cite{Easther:2006vd,GarciaBellido:2007dg,GarciaBellido:2007af,Dufaux:2007pt}, first order cosmological phase transitions \cite{Witten:1984rs,Hogan:1986qda}, topological defects \cite{Durrer:1998rw,Bevis:2006mj,Leblond:2009fq,Hiramatsu:2010yz}, braneworlds \cite{Randall:2006py,Durrer:2007ww}.
These sources have their own characteristic spectral patterns.
So, it may be possible to distinguish a physical phenomenon contributing to a signal of GWB which may be a mixture of various contributions.    

Meanwhile, the thermal history of the universe could have critical impact on GWBs.
For example, if there were an early matter-domination (MD) era, the spectrum of a GWB generated during or before that era could be very different from what is expected in the absence of such an era \cite{Seto:2003kc,Nakayama:2008ip,Nakayama:2008wy,Kuroyanagi:2011fy,Kuroyanagi:2013ns,Jinno:2014qka}.
In this case, generically the amplitude of a signal is reduced, depending on the duration of the matter-domination era and frequencies.
Hence, it is possible to see no GWB signal in direct detection experiments due to a significant reduction of the initial amplitude.
In this sense, the lower the energy scale of GWB generation is, the more likely to be safe from a possible reduction a signal is, having more chances to be detected.

A GWB from a first order phase transition associated with weak or $\TeV$ scale physics of either extensions of standard model (SM) or hidden sector is well motivated and may be detectable in future experiments such as eLISA \cite{eLISA} or LISA \cite{LISA} (see for example Ref.~\cite{Caprini:2015zlo} and references therein).
It has been studied for many years \cite{Witten:1984rs,Hogan:1986qda,Turner:1990rc,Kosowsky:1992rz,Kosowsky:1991ua,Turner:1992tz,Kosowsky:1992vn,Kamionkowski:1993fg,Apreda:2001us,Grojean:2006bp,Caprini:2006jb,Caprini:2007xq,Easther:2008sx,Caprini:2009fx,Caprini:2009yp,Megevand:2009gh,Espinosa:2010hh,Hindmarsh:2013xza,Caprini:2015zlo,Hindmarsh:2015qta,Huang:2016odd}, and is recently attracting attention again \cite{Jaeckel:2016jlh,Dev:2016feu,Hashino:2016rvx}, stimulated by the recent great discovery of blackhole-sourced gravitational waves at LIGO \cite{LIGO,Abbott:2016blz}.  
If such a signal is sensed at any detector, it may be very useful to probe the early universe of energy around $\TeV$ scale.
Especially, it might be possible to know if there was a MD era which ended after the generation of such a GWB, and hence the decay rate of the dominating matter would be constrained.
If we can find it, such a constraint would have direct relevance, for example, on baryogenesis scenarios and the symmetry breaking scale of supersymmetry (or the mass scale of Planckian moduli which is expected to have Planck-supressed interactions to visible sector particles \cite{Coughlan:1983ci,Banks:1993en,deCarlos:1993wie}) .
However, this intriguing possibility or benefit of having a GWB from a first order phase transition depends on whether we can recover the governing parameters from a signal or not.

In this work, we first investigate a stochastic GWB from a short-lasting source in a MD era, showing that, as it is expected, the spectrum of the GWB right after the generation epoch is the same as the one generated in  a radiation-domination (RD) era. 
Subsequently, we discuss the spectrum of a GWB from a first order phase transition in a MD era, showing that there appears a unique characteristic feature which allows to determine when the phase transition took place and the matter-domination era ended.
Then, we discuss degeneracies in a GWB signal, showing that parameters can span a wide range covering a couple of orders of magnitude for a specific amplitude and shape of the signal, although the degeneracies depend on detector sensitivities.


\section{A stochastic background of gravitational waves}
In a flat FLRW universe, the tensor metric perturbation is defined as
\beq
ds^2 = a^2(\eta) \l[ d\eta^2 - \l( \delta_{ij} + h_{ij} \r) dx^i dx^j \r]
\eeq 
with $a$ and $\eta$ being the scale factor and conformal time, respectively.
The energy density of the gravitational wave (GW) is defined as 
\beq
\rho_{\rm GW} \equiv \frac{\langle \dot{h}_{ij}(\bold{x},\eta) \dot{h}^{ij}(\bold{x},\eta) \rangle}{32 \pi G a^2}
\eeq
where ``$\ \dot{} \ $'' denotes a derivative w.r.t $\eta$.
Then, defining a spectral density as 
\beq
\langle \dot{h}_{ij}(\bold{k},\eta) \dot{h}^{ij}(\bold{q},\eta) \rangle \equiv (2 \pi)^3 \delta^3(\bold{k}+\bold{q}) P_{\dot{h}}(\bold{k},\eta),
\eeq
the conventional (normalized) energy density of GWs per logarithmic interval of comoving wave number is found to be
\beq
\Omega_{\rm GW} \equiv \frac{1}{\rho_{\rm c, 0}} \frac{d \rho_{\rm GW}}{d \ln k} = \frac{k^3 P_{\dot{h}}(\bold{k},\eta)}{8 (2 \pi)^3 G \rho_{\rm c,0} a^2} 
\eeq
where $\rho_{\rm c,0}$ is the critical energy density at present.

The sources of GWs are transverse-traceless parts of tensor-type anisotropic stresses, denoted here as $\Pi_{ij}$.
For a given source, in Fourier space the linearized Einstein field equation is given by
\beq \label{h-eom-eta}
\ddot{h}_{ij} + 2 \mathcal{H} \dot{h}_{ij} + k^2 h_{ij} = 16 \pi G a^2 \Pi_{ij}
\eeq
where $\mathcal{H} \equiv d\ln a/d\eta$.
The evolution equations of the source are obtained from the conservation equations of the energy momentum tensor, $T^{\mu\nu}_{\phantom{\mu\nu};\nu}=0$.
It has been shown that for fluids of relativistic gas in an expanding flat universe dominated by radiation, the dynamical variables can be rescaled by appropriate powers of the scale factor such that the evolution equations of the rescaled energy and momentum densities of fluids in conformal time have the same form as the ones in a flat non-expanding universe \cite{Brandenburg:1996fc}.  
Hence, GWs in an expanding universe can be easily mapped out from ones in Minkowsky spacetime if the source is of the same type and can be rescaled properly.

In a matter-dominated universe, the form of the fluid equations is not maintained due to the explicit time dependence of the decay of the dominating matter, and there is a continuous energy injection to radiation from the decay.
In addition, the radiation background and the scalar field going through a phase transition can interact with the background matter.
Hence, strictly speaking, in order to get a spectrum of GWs generated during MD era one should solve the fluid equations directly.
However, when the dominating matter decays well after the phase transition, the very small decay rate implies that the scattering between radiation/scalar and matter can be negligible during the generation of a GWB from the phase transition.
Also, if a source of GWs is active only for very short time period $\Delta t$ such that $H \Delta t \ll 1$ with $H$ being the expansion rate defined w.r.t cosmic time $t$ around the epoch of the generation of GWs, the energy injection to radiation can be ignored.
In this case, the fluid equations can be properly rescaled in the same way as the case of the radiation-dominated universe, and we can utilize the known results of a radiation-dominated universe (see Ref.~\cite{Caprini:2015zlo} for example) with appropriate additional red-shift effect taken into account.  
From now on, we consider this simple case (short-lasting source and negligible change of radiation background during the life time of a source).
The case of long-lasting sources will be studied elsewhere.

The source can be expressed as
\beq \label{source}
\Pi_{ij}(k,\eta) \equiv \l( \rho_{\rm s} + p_{\rm s} \r) \tilde{\Pi}_{ij}(k,\eta) = \frac{1}{2 \pi G} \l( \frac{\rho_{\rm s}}{\rho_{\rm c}} \r) \frac{\mathcal{H}^2}{a^2} \tilde{\Pi}_{ij}(k,\eta)
\eeq
where $\rho_{\rm s}$ and $p_{\rm s} (= \rho_s/3)$ are respectively the energy  and pressure densities of the source fluid, and $\rho_{\rm c}$ is the critical energy density at a given conformal time $\eta$.
Note that under our assumption (of negligible energy injection to radiation) $\rho_{\rm s} \propto a^{-4}$ during the active time of the source and the dimensionless quantity $\tilde{\Pi}_{ij}$ is expected not to depend on the expansion history of the universe as discussed in the previous paragraph.
Hence, a spectral pattern derived from $\tilde{\Pi}_{ij}$ in a radiation-dominated or non-expanding universe can be applicable to the case of matter-dominated universe for a given source. 

In an expanding universe dominated by matter, the scale factor well after a reference point of $a_{\rm ref}=1$ is given by
\beq
a \simeq \frac{1}{4} H_{\rm ref}^2 \eta^2
\eeq
where $H_{\rm ref}$ is the Hubble parameter at the reference point.
Hence, introducing a new variable $\tilde{\eta} \equiv k \eta$ for convenience, we can rewrite \eq{h-eom-eta} as
\beq \label{h-eom-x}
h_{ij}''+\frac{4}{\tilde{\eta}} h_{ij}' + h_{ij} = 32 \l( \frac{\rho_{\rm s}}{\rho_{\rm c}} \r)_{\rm in} \l( \frac{\tilde{\eta}_{\rm in}}{\tilde{\eta}} \r)^2 \frac{\tilde{\Pi}_{ij}}{\tilde{\eta}^2}
\eeq
where `$\prime$'$=d / d\tilde{\eta}$ and subscript `$_{\rm in}$' denotes the time when the souce start being active.
The solution of \eq{h-eom-x} is 
\beq \label{h-sol}
h_{ij}(k,\tilde{\eta}) = 32 \l( \frac{\rho_{\rm s}}{\rho_{\rm c}} \r)_{\rm in} \tilde{\eta}_{\rm in}^2 \int_{\tilde{\eta}_{\rm in}}^{\tilde{\eta}} d\tilde{\eta}' G(\tilde{\eta},\tilde{\eta}') \frac{\tilde{\Pi}_{ij}(k,\tilde{\eta}'])}{\tilde{\eta}'^4}
\eeq
and 
\beq
G(\tilde{\eta},\tilde{\eta}') = \frac{\tilde{\eta}'}{\tilde{\eta}^3} \l[ \l( \tilde{\eta} \tilde{\eta}' +1 \r) \sin(\tilde{\eta}-\tilde{\eta}') + \l( \tilde{\eta}-\tilde{\eta}' \r) \cos(\tilde{\eta}-\tilde{\eta}') \r]
\eeq
Hence, for $1 \ll \tilde{\eta}' < \tilde{\eta}$, we find
\beq
h_{ij}'(k,\tilde{\eta}) \simeq 32 \l( \frac{\rho_{\rm s}}{\rho_{\rm c}} \r)_{\rm in} \tilde{\eta}_{\rm in}^2 \int_{\tilde{\eta}_{\rm in}}^{\tilde{\eta}} d\tilde{\eta}' \cos(\tilde{\eta}-\tilde{\eta}') \frac{\tilde{\Pi}_{ij}(k,\tilde{\eta}')}{\tilde{\eta}^2 \tilde{\eta}'^2}
\eeq
giving 
\begin{widetext}
\bea \label{P-dh}
P_{h'} 
&=& \frac{1}{2} \l[ 32 \l( \frac{\rho_{\rm s}}{\rho_{\rm c}} \r)_{\rm in} \frac{\tilde{\eta}_{\rm in}^2}{\tilde{\eta}^2} \r]^2 \int_{\tilde{\eta}_{\rm in}}^{\tilde{\eta}} d\tilde{\eta}_1 d\tilde{\eta}_2 \frac{\cos(\tilde{\eta}_2-\tilde{\eta}_1)}{\tilde{\eta}_1^2 \tilde{\eta}_2^2} \tilde{\Pi}(k,\tilde{\eta}_1,\tilde{\eta}_2)
\nonumber \\
&\simeq& \frac{1}{2} \l[ 8 \l( \frac{\rho_{\rm s}}{\rho_{\rm c}} \r)_{\rm in} \r]^2 \frac{\mathcal{H}^4(\eta)}{k^4} \int_{\tilde{\eta}_{\rm in}}^{\tilde{\eta}} d\tilde{\eta}_1 d\tilde{\eta}_2 \cos(\tilde{\eta}_2-\tilde{\eta}_1) \tilde{\Pi}(k,\tilde{\eta}_1,\tilde{\eta}_2)
\eea
\end{widetext}
where the unequal-time correlator is defined as \cite{Caprini:2009yp}
\beq
\langle \tilde{\Pi}_{ij}(k,\tilde{\eta}_1) \tilde{\Pi}^{ij}(k,\tilde{\eta}_2) \rangle \equiv (2 \pi)^3 \delta^3(\bold{k}-\bold{q}) \tilde{\Pi}(k,\tilde{\eta}_1,\tilde{\eta}_2)
\eeq
and in the second line $\tilde{\eta}_{1,2} \approx \tilde{\eta}_{\rm in}$ was used that is relevant for short-lasting sources.
Comparing \eq{P-dh} to the case of radiation-domination (RD) (see for example \cite{Caprini:2009yp}), we find that for a short-lasting source the spectrum of a GWB right after its generation in MD is the same as that in RD modulo the extra factor $(\rho_{\rm s}/\rho_{\rm c} )_{\rm in}^2$.
The GW amplitude after the generation epoch is found to be 
\begin{widetext}
\beq
\Omega_{\rm GW}(\bold{k},\eta) 
= \l( \frac{\rho_{\rm s}}{\rho_{\rm c}} \r)_{\rm in}^2 \frac{1}{a^4} \frac{4 H_{\rm in}^4 k}{3 \pi^2 H_0^2} \int_{\tilde{\eta}_{\rm in}}^{\tilde{\eta}} d\tilde{\eta}_1 d\tilde{\eta}_2 \cos(\tilde{\eta}_2-\tilde{\eta}_1) \tilde{\Pi}_{ij}(k,\tilde{\eta}_1,\tilde{\eta}_2)
\eeq
\end{widetext}
where we set $a_{\rm ref} = a_{\rm in}=1$.
%
%
%
From a numerical fitting, we find that the expansion rate as a function of the scale factor can be very well approximated as
\beq \label{H-a}
H(a) = H_{\rm ref} \l( \frac{a_{\rm ref}}{a_\times} \r)^{\frac{3}{2}} \l( \frac{a_\times}{a} \r)^2 \l[ 1+\l(\frac{a_\times}{a} \r)^2 \r]^{-\frac{1}{4}}
\eeq
across the region of the matter-radiation transition with errors less than $5$ \%, and $a_{\rm ref}/a_\times \equiv (7/9) \Gamma_{\rm d}/H_{\rm ref}$ with $\Gamma_{\rm d}$ being the decay rate of the dominating matter.
From \eq{H-a}, the scale factor can be expressed as 
\begin{widetext}
\beq \label{red-shift}
\l( \frac{a}{a_\times} \r)_{\rm MD}
= \l[ \frac{\sqrt{2} \l( g_*(H)/g_*(H_\times) \r)_{\rm RD}^{\frac{1}{3}}}{ \sqrt{1 + \l( a_\times/a \r)_{\rm MD}^2}} \r]^{\frac{1}{4}} \l( \frac{a}{a_\times} \r)_{\rm RD}
= \l[ \sqrt{2 \l( \frac{k_\times}{k} \r)^4+\frac{1}{4}} - \frac{1}{2} \r]^{1/2}
\eeq
\end{widetext}
where $g_*(H)$ is the expected number of relativistic degrees of freedom in a radiation-dominated universe with the expansion rate $H$ \footnote{We set $g_*(H_*)=100$ for numerical analysis irrespective of $H_*$ (or $T_*$).}, and the last equality is for $k\geq k_\times$ as the comoving wave numbers at the horizon crossing. 
From now on, we use a subscript ``$_*$'' for the quantities at the epoch of the generation of GWs from a short-living source.
Then, 
the present amplitude of GWs can be expressed as
\begin{widetext}
\beq \label{Omega-GW-0}
\frac{\Omega_{GW}^{\rm MD}(k,\eta_0)}{\l( \rho_{\rm s}/\rho_{\rm c} \r)^2_* \Omega_{GW}^{\rm RD}(k,\eta_0)}  
\simeq  
\mathcal{F}^4(k,k_\times,k_*) \equiv 
\l\{ \begin{array}{lcc}
\sqrt{2} \l( \frac{g_*(H_*)}{g_*(H_\times)} \r)_{\rm RD}^{\frac{1}{3}} \l[ 1+ \l( \frac{H_*}{2^{1/4} H_\times} \r)^{\frac{4}{3}} \r]^{-\frac{1}{2}}  & {\rm for} & k>k_*
\\
\sqrt{2} \l( \frac{g_*(H)}{g_*(H_\times)} \r)_{\rm RD}^{\frac{1}{3}} \l[ \frac{\sqrt{2 \l( \frac{k_\times}{k} \r)^4 + \frac{1}{4}} - \frac{1}{2}}{\sqrt{2 \l( \frac{k_\times}{k} \r)^4 + \frac{1}{4}} + \frac{1}{2}} \r]^{\frac{1}{2}} & {\rm for} & k_\times < k \leq k_*
\\
1 & {\rm for} & k \leq k_\times
\end{array}
\r.
\eeq
\end{widetext}
where $H_\times = 2^{-1/4}(7/9) \Gamma_{\rm d}$, and $\Omega_{GW}^{\rm RD}(k,\eta_0)$ is understood to take into account the changes of characteristic quantities in MD relative to the ones in RD as well as the additional redshifts of frequencies (see the next section).

\section{Gravitational waves from a first order phase transition in MD}

\begin{figure}[t]
\begin{center}
\includegraphics[width=0.48\textwidth]{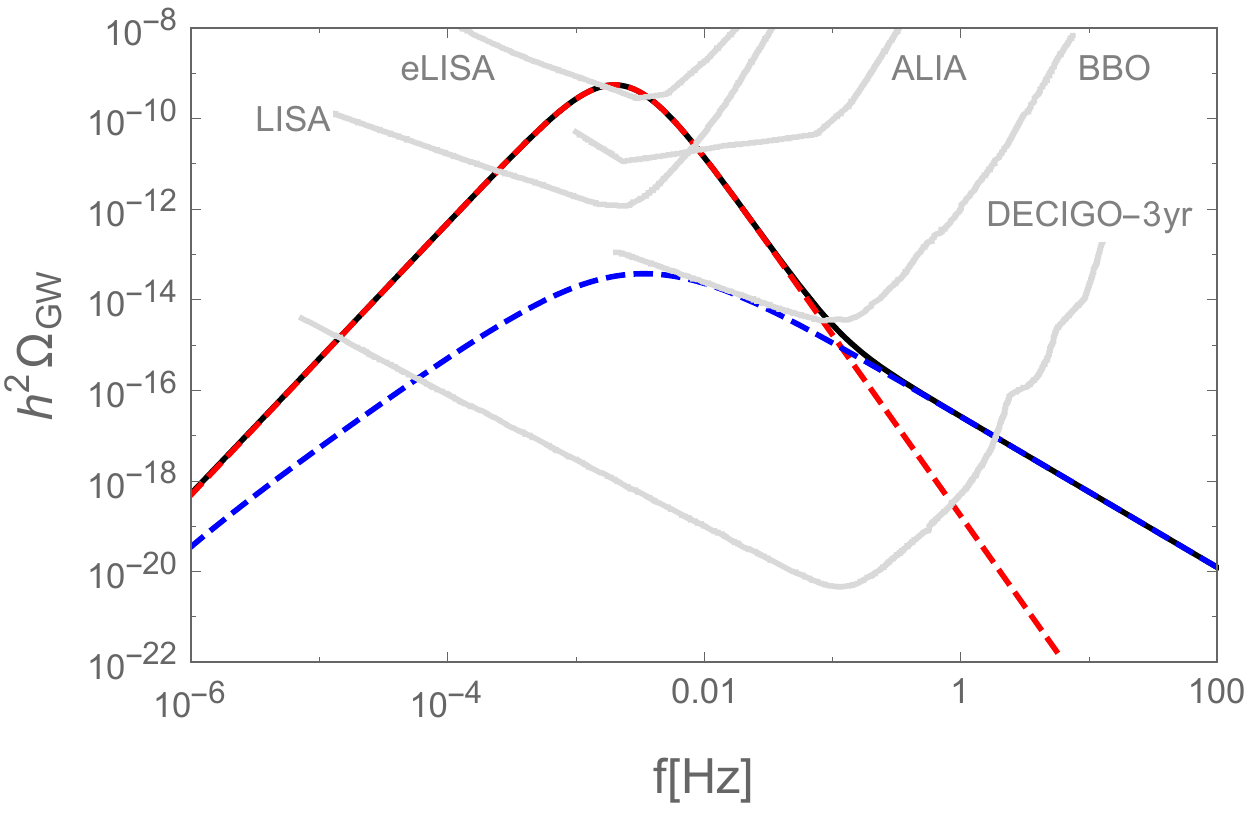}
\includegraphics[width=0.48\textwidth]{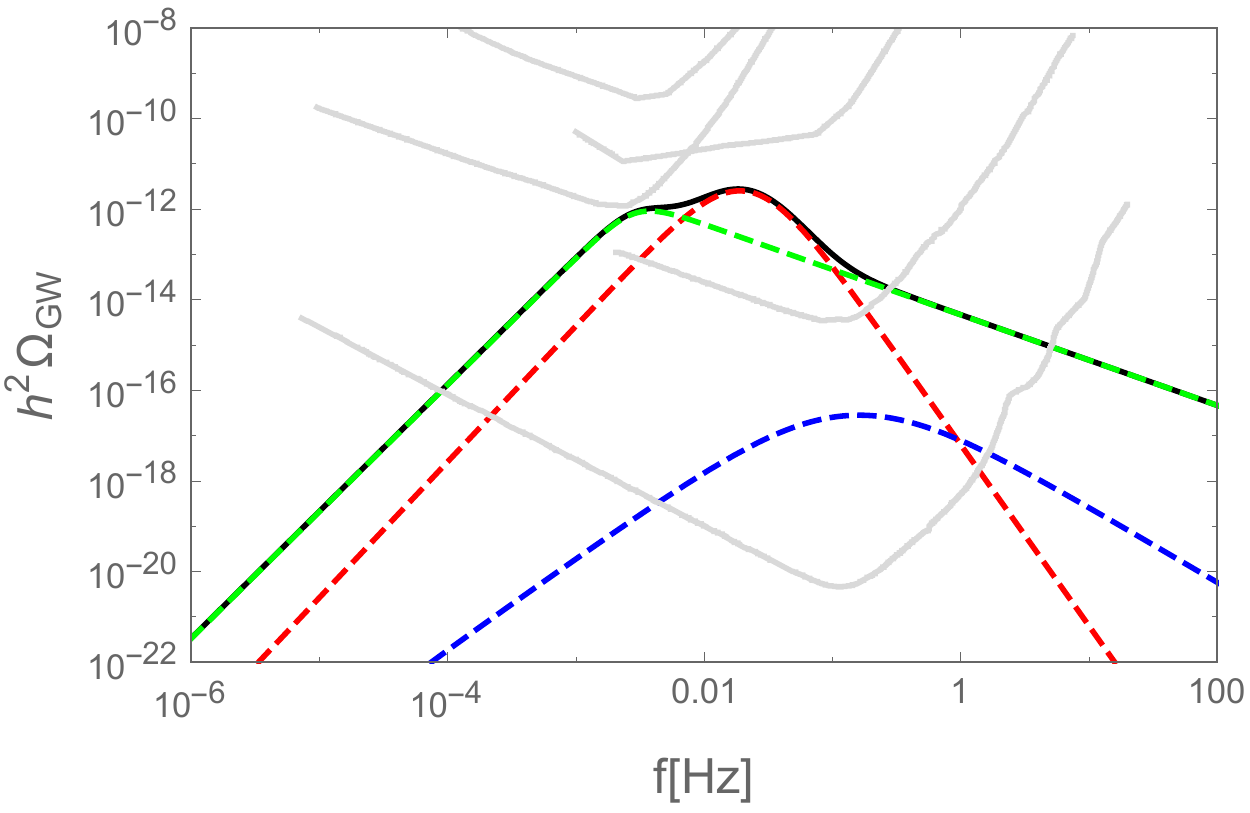}
\caption{GWs in RD. 
\textit{Top}: Non-runaway bubbles with $\alpha=0.5, \ \beta/H_*=100, \ v_w=0.95, \ T_*=100 \GeV$.
Red and blue dashed lines are contributions from sound waves and turbulence, respectively.
Black solid line is the sum of those two contributions.
\textit{Bottom}: Runaway bubbles in Plasma with $\alpha_\infty=0.1, \ \alpha=0.2, \ \beta/H_*=100, \ v_w=1, \ T_*=1 \TeV$. 
Color scheme is the same as Top panel except the green line which represents the contribution from bubble collisions.
}
\label{fig:GW-RD}
\end{center}
\end{figure}

Gravitational waves from a first order phase transition in RD have been studied for many years, and the expected spectra from various contributions are now rather well known although numerical simulations should be improved to cover a wider range of parameter space (see for example Ref.~\cite{Caprini:2015zlo} and references therein).
For convenience, the estimations of GWs found in literature were collected in an Appendix, and Fig.~\ref{fig:GW-RD} shows the expected signals in RD for a set of parameters for each bubble dynamics.
In the figure, $\epsilon=0.05$ was used as the ratio of energy efficiency factors $\kappa_{\rm turb}/\kappa_{\rm sw}$, and it will be used through out this paper for numerical analysis.
Among the three sources (bubble collisions, sound waves and turbulence) of gravitational waves from a first order phase transition, the sound waves and turbulence are long-lasting sources having lifetimes of Hubble scale or longer.
However, as shown in Ref.~\cite{Caprini:2009yp}, the contribution of the long-lasting part of turbulence is only a slight overall increase (by about a factor of two) of the amplitude, barely changing the spectrum as compared to the case of only short-lasting part taken into account.
This may be true even in the case of sound waves since the slope of the spectrum in the low frequency region is steeper than the case of turbulence.
Motivated by this result, we will regard all those sources as short-lasting ones in the following discussion as long as $\beta/H_* \gtrsim \mathcal{O}(10-100)$. 

In matter domination, the nature of a phase transition is expected to be changed, as follows, due to the more rapid expansion rate relative to the case of RD.
In a first order phase transition, the bubble nucleation rate per unit volume and time is given by \cite{Linde:1978px,Linde:1981zj,Hogan:1984hx}
\beq
\Gamma(T) \sim T^4 e^{-S_3(T)/T}
\eeq
where $S_3(T)$ is the three-dimensional Euclidean action of the bounce solution associated with the scalar field of interest.
The dominant bubbles at percolation are nucleated at \cite{Easther:2008sx}
\beq
t_n = t_p - \frac{3}{\beta(t_n)}
\eeq
where $t_p$ is the time when the phase transition ends, and $\beta \equiv d \ln \Gamma / d t$.
The nucleation rate at $t_n$ is estimated as
\beq
\Gamma(t_n) = \frac{\beta(t_n)^3}{8\pi e^3}
\eeq
and the temperature at the time is found from
\beq \label{Tn}
\exp\l[-\frac{S_3}{T_n} \r] = \frac{1}{8 \pi e^3} \l( \frac{\beta}{H} \r)^4 \l( \frac{H}{T_n} \r)^4 
\eeq
Note that $H \propto T^4 \propto a^{-3/2}$ in MD, leading to 
\beq
\beta(T)_{\rm MD} = \frac{3}{8} \beta(T)_{\rm RD}
\eeq
and  
\beq
H(T)_{\rm MD} \simeq \frac{5}{2} \frac{H^2(T)_{\rm RD}}{\Gamma_{\rm d}}
\eeq 
Hence,
\beq \label{beta-MD}
\l( \frac{\beta(T)}{H(T)} \r)_{\rm MD} = \frac{3}{20} \l( \frac{\Gamma_\phi}{H(T)} \r)_{\rm RD} \l( \frac{\beta(T)}{H(T)} \r)_{\rm RD}
\eeq
and \eq{Tn} can be rewritten as
\beq \label{Tn-MD}
\l. \exp\l[-\frac{S_3}{T_n} \r] \r|_{\rm MD} = \frac{(3/8)^4}{8 \pi e^3} \l[ \l( \frac{\beta}{H} \r)^4 \l( \frac{H}{T_n} \r)^4 \r]_{\rm RD}  
\eeq
\eq{Tn-MD} implies that $T_n$ in MD should be slightly higher than the one expected in RD, requiring $(S_3/T_n)_{\rm MD} - (S_3/T_n)_{\rm RD} \simeq 4$.
From the definition of $\beta$, one finds
\beq
\frac{\Delta T}{T} \simeq \frac{\Delta (S_3/T)}{\beta/H}
\eeq
Hence, if $\beta(T_n)/H(T_n) \gg 4$, the change of $T_n$ in MD relative to the case of RD can be ignored.
In order to make  a clear comparison to the case of RD, we consider only this case in the subsequent discussion.

\begin{figure}[t]
\begin{center}
\includegraphics[width=0.48\textwidth]{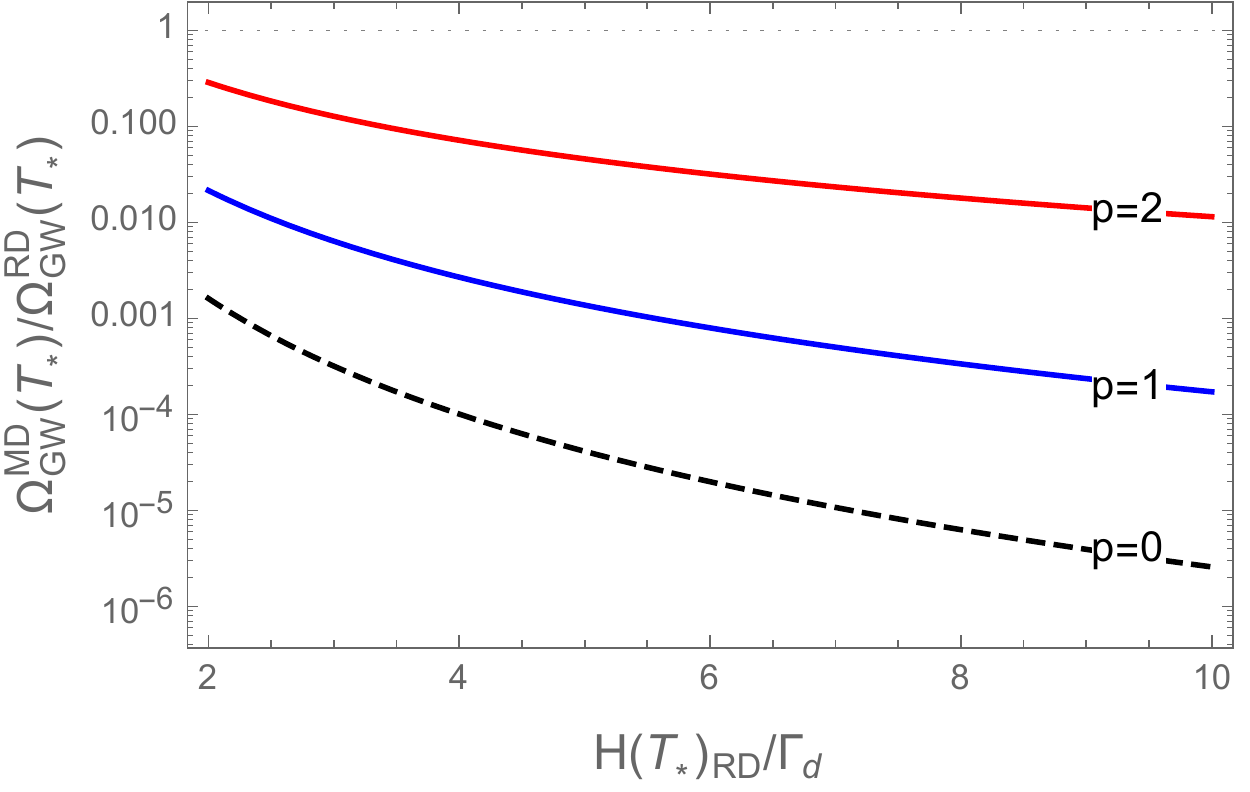}
\caption{$\Omega_{\rm GW}^{\rm MD}/\Omega_{\rm GW}^{\rm RD}$ in \eq{Omega-MD-to-RD} for a temperature at the peak frequency of each contribution.
}
\label{fig:MD-to-RD}
\end{center}
\end{figure}
\begin{figure}[h!]
\begin{center}
\includegraphics[width=0.48\textwidth]{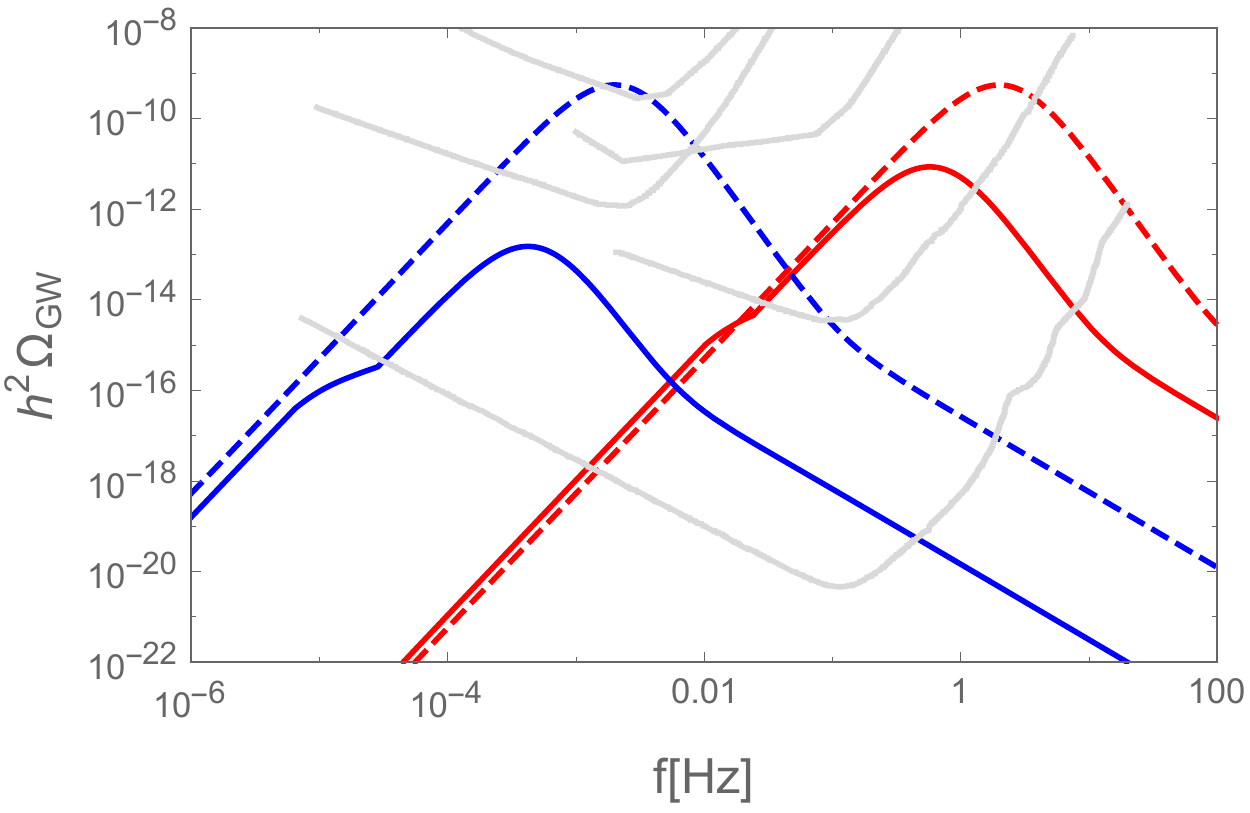}
\includegraphics[width=0.48\textwidth]{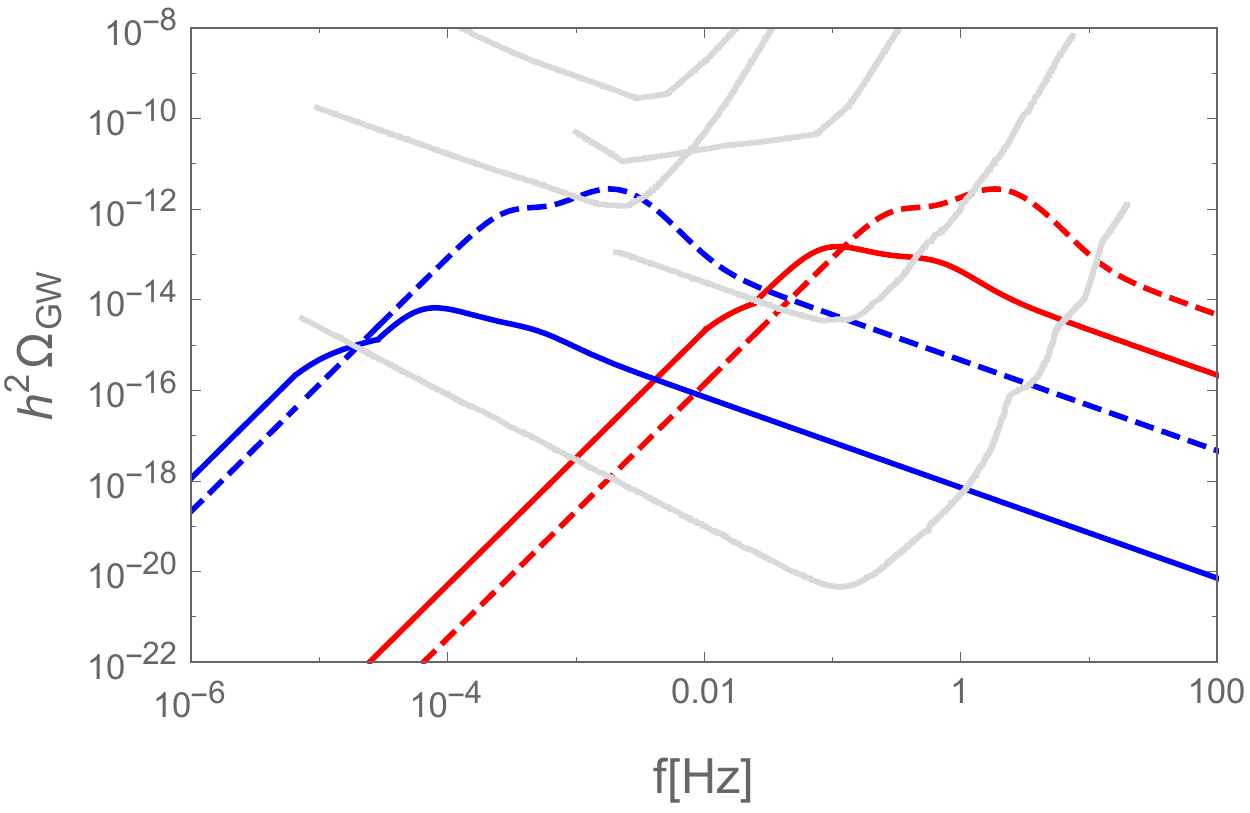}
\caption{GWs in MD (solid lines) relative to the ones in RD (dashed lines) for the same $T_*$ as in MD.
.
Red lines are for $(T_*,T_{\rm d})=(10^5 \GeV, 8 \times 10^4 \GeV)$.
Blue lines are for $(T_*, T_{\rm d})=(100 \GeV, 50 \GeV)$.
\textit{Top}: Non-runaway bubbles with $\alpha=0.5, \ (\beta/H_*)_{\rm RD} = 100, \ v_w = 0.95$.
\textit{Bottom}: Runaway bubbles in Plasma with $\alpha_\infty = 0.1, \ \alpha=0.2, \ (\beta/H_*)_{\rm RD}=100, \ v_w=1$.
}
\label{fig:GW-spec-MD}
\end{center}
\end{figure}
As one can see from the estimation of GWs in RD (see Appendix), GWs can be regarded as a function of 
\beq
\alpha, \ \beta, \ v_w \ ({\rm or} \ \alpha_\infty), \ H_*
\eeq  
modulo the spectral behaviors encoded in $S_i(f)$ which are expected to be same for both RD and MD in our consideration.
The bubble wall velocity is determined by the friction parameter and the ratio of latent heat to the radiation density at the symmetric phase ($\alpha$) \cite{Espinosa:2010hh}.
Both of them are temperature-dependent.
Hence, if the nucleation temperature $T_n$ is nearly the same in both RD and MD, so are $\alpha(T_n)$ and $v_w(T_n)$.
In this case, the main difference of MD and RD in terms of the peak amplitude of $\Omega_{\rm GW}$ is from the dependence on $\beta/H_*$.
Each contribution of GWs sources in \eqss{Omega-b}{Omega-sw}{Omega-turb} is proportional to $(H_*/\beta)^p$ with $p=1,2$ depending on sources.
Hence, from \eqs{Omega-GW-0}{beta-MD}, for the peak frequency in each case of RD and MD one finds
\beq \label{Omega-MD-to-RD}
\frac{\Omega_{\rm GW}^{\rm MD}(T_*)}{\Omega_{\rm GW}^{\rm RD}(T_*)} \simeq \l( \frac{2}{5} \frac{\Gamma_{\rm d}}{H(T_*)_{\rm RD}} \r)^4 \l[ \frac{20}{3} \l( \frac{H(T_*)_{\rm RD}}{\Gamma_{\rm d}} \r) \r]^p  
\eeq
where $\Omega_{\rm GW}^{RD}(T_*)$ is the amplitude of  GWs generated at $T \approx T_*$ with a $\beta(T_n \simeq T_*)$ in RD.
Fig.~\ref{fig:MD-to-RD} shows $\l(\Omega_{\rm GW}^{\rm MD}/\Omega_{\rm GW}^{\rm RD}\r)_{T=T_*}$ as a function of $H(T_*)_{\rm RD}/\Gamma_{\rm d}$ and $p$.
The case of $p=0$ represents when only an overall suppression of the energy and momentum densities of a given source relative to the matter density is taken into account.
As shown in the figure, the ratio $\l(\Omega_{\rm GW}^{\rm MD}/\Omega_{\rm GW}^{\rm RD}\r)_{T=T_*}$ in the cases of $p\neq0$ is larger than the one with $p=0$ at least by one or two orders of magnitude.
This presents a good perspective in terms of detection, but may cause more degeneracy as will be discussed in the next section. 

The characteristic frequencies of \eqss{f-b}{f-sw}{f-turb} are also modified in MD due to the changes in $\beta/H_*$ and the relation between $T_*$ and $H_*$.
Since the present values are obtained as 
\beq
f_i 
\propto \l( \frac{\beta}{v_w} \r) \l( \frac{a_*}{a_\times} \r) \l( \frac{a_\times}{a_0} \r),
\eeq
from \eqs{red-shift}{beta-MD}, their values in MD and RD are related as
\beq
f_i^{\rm MD}(T_*) = \frac{3}{8} \mathcal{F}(k>k_*) \times f_i^{\rm RD}(T_*) 
\eeq 
where we assumed that $v_w$ and $f_*/\beta$ in \eq{fstar-beta} are not changed as long as the temperature at the time of the phase transition is nearly unchanged in MD relative to that of RD.  

Fig.~\ref{fig:GW-spec-MD} shows expected GW spectra in MD for a couple of parameter sets as examples.
In the figure, a kink-like change of spectrum appears at the low-frequency side of the peak position, which is caused by the mode-dependent redshift described in \eq{Omega-GW-0}.
Also, for lower frequencies the spectrum merges to the one expected in the case of radiation-dominated universe.
The ratio of the peak frequency to the one at the kink-like change is nothing but 
\beq \label{f-beta}
\frac{f_{\rm peak}}{f_{\rm kink-like}} = 
\l\{
\begin{array}{ll}
(f_*/\beta) (\beta/H_*) & {\rm for \ collisions} \\
(\beta/H_*)/v_w & {\rm for \ others}
\end{array}
\r.
\eeq  
The ratio of the kink-like change to the merging point is 
\beq \label{f-Td}
\frac{f_{\rm kink-like}}{f_{\rm merger}} = \frac{a_* H_*}{a_\times H_\times} = 2^{\frac{1}{6}} \l( \frac{H_*}{H_\times} \r)^{\frac{1}{3}} = 2^{\frac{1}{4}} \l( \frac{9 H_*}{7 \Gamma_{\rm d}} \r)^{\frac{1}{3}}
\eeq
The smaller $(\beta/H_*)/v_w$ is, the closer to the peak the kink-like change becomes, allowing easy detection.


\section{Probing the early universe by gravitational waves}
In a radiation-dominated universe eight parameters are needed to determine $h^2 \Omega_{\rm GW}$ generated from a first order phase transition (see Appendix):
\beq
\alpha, \ \frac{\beta}{H_*}, \ \kappa_b, \ \kappa_{\rm sw}, \ \kappa_{\rm turb}, \ v_{\rm w} ({\rm or} \ \alpha_\infty), \ H_*
\eeq
However, as shown in Ref.~\cite{Espinosa:2010hh}, $v_{\rm w}$ is nearly determined as a function of $\alpha$ and the  friction parameter coming from microphysics (see also Ref.~\cite{Megevand:2009gh}).
Hence, either the friction parameter or $v_{\rm w}$ can be regarded as a free parameter, depending on its relevance.
Then, all $\kappa_i$s can be regarded as functions of $\alpha$ and $v_{\rm w}$ or $\alpha_\infty$ which is directly obtained from the friction parameter.
Therefore, practically the amplitude and shape of GWs are determined only by four parameters:
\beq \label{free-para}
\alpha, \ \frac{\beta}{H_*}, \ v_{\rm w} ({\rm or} \ \alpha_\infty), H_*
\eeq

On the other hand, when GWs are measured in future experiments, the information contained in the signal depends on the spectral shape (e.g., peak frequency and amplitude, slope, and any extra features).
In the case of GWs from a first order phase transition, there can be two or three contributions (see Appendix), depending on whether a terminal velocity of bubble wall exists or not. 
For a given set of parameters in \eq{free-para}, each contribution has a specific amplitude and frequency at the peak, and a specific slope of its spectrum.
For a given contribution, only two pieces of information (peak frequency and amplitude) are relevant while there are four free-parameters.
The peak (or characteristic) frequency $f_i$ in \eqss{f-b}{f-sw}{f-turb} is determined by $(\beta/H_*) T_*/v_w$.
Since $(\beta/H_*)/v_w >1$, one finds
\beq \label{Tstar-ubnd}
\frac{f_i/\gamma_i}{{\rm mHz}} \l( \frac{v_w}{\beta/H_*} \r) \leq \frac{T_*}{1 \TeV} \l( \frac{g_*(T_*)}{100} \r)^{\frac{1}{6}} < \frac{f_i/\gamma_i}{{\rm mHz}}
\eeq
where $\gamma_i=(0.038,0.19,0.27)$ for the contributions of bubble collisions, sound wave, and turbulence, respectively.
That is, once a GW signal is measured with a peak structure, $T_*$ can be upper-bounded but can not be fixed yet.
The $(\beta/H_*)/v_w$-dependence in the peak amplitude can be traded off with the peak frequency and $T_*$.
Then, for a peak frequency, the peak amplitude is determined by a combination of $T_*, \alpha, v_w (\ {\rm or} \ \alpha_\infty)$.
Hence, still  two parameters remain undetermined.
If the role of main contribution is sensitively changed as compared to the other contributions, it will imply that the appearance of a  specific contribution may add another piece of information to constrain the range of parameters.
However this happens in a limited region of parameter space (and mostly for runaway bubbles in plasma).
This implies that, if a future expected GW signal matches well with a single contribution out of  the three potential sources of a first order phase transition, there will be degeneracies among macroscopic parameters and it would be difficult to know what microphysical model may be responsible for the signal.
%
\begin{figure}[t]
\begin{center}
\includegraphics[width=0.48\textwidth]{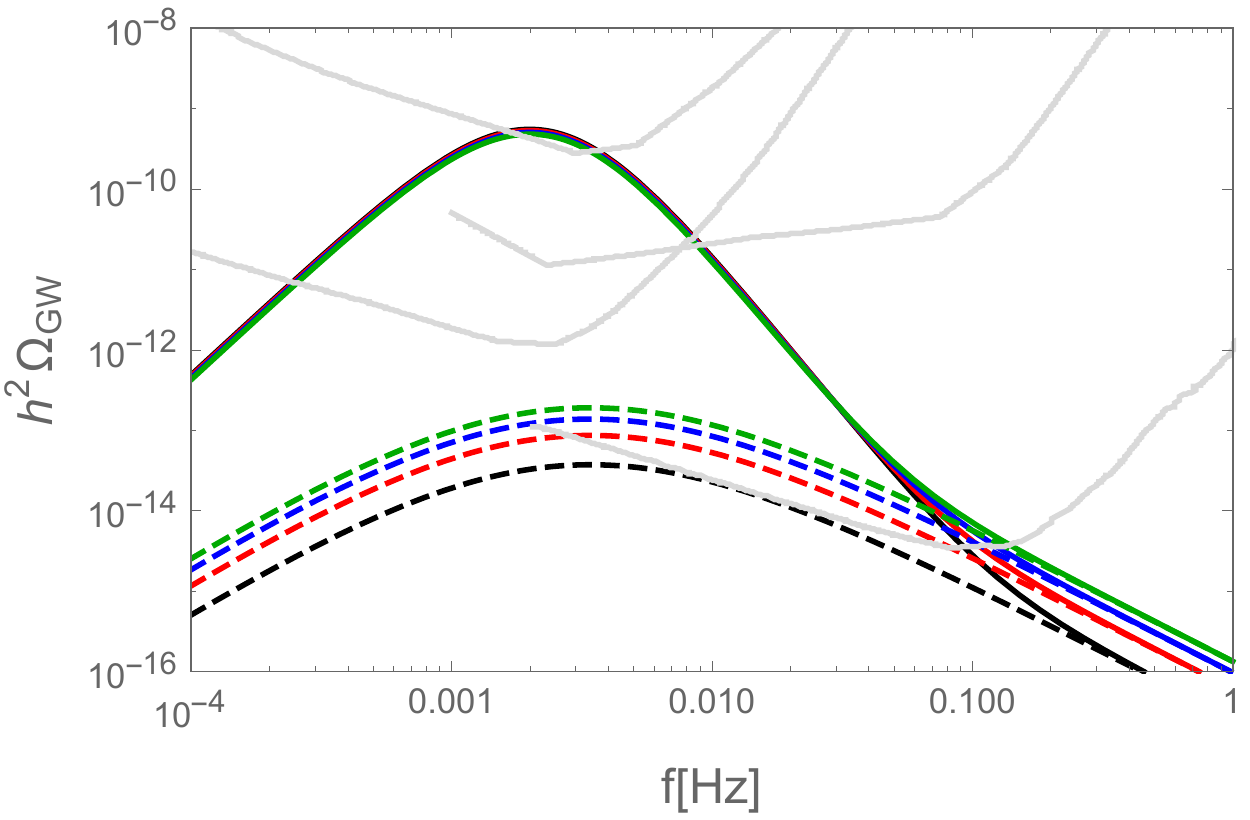}
\caption{Degeneracy (solid lines) in GW spectra in RD for non-runaway bubbles with different choices of parameters. 
Solid lines are summations of the contributions from sound waves and turbulences.
Each dashed line is the contribution of turbulence only for a specific set of parameters.
We took $v=0.95$ for all different lines.
For black lines, $\alpha = 0.5$ was taken.
For the each of other color lines, $\alpha$ was reduced by a factor $2^{2.5},3^{2.5},4^{2.5}$ for red, blue, and green (solid and dashed) lines, respectively.
Also, the same factors were applied to $\beta/H_*$ but the inverse of the factor applied to $T_*$ so as to keep $(\beta/H_*)T_*$ fixed.
}
\label{fig:GW-RD-deg1}
\end{center}
\end{figure}
\begin{figure}[t]
\begin{center}
\includegraphics[width=0.48\textwidth]{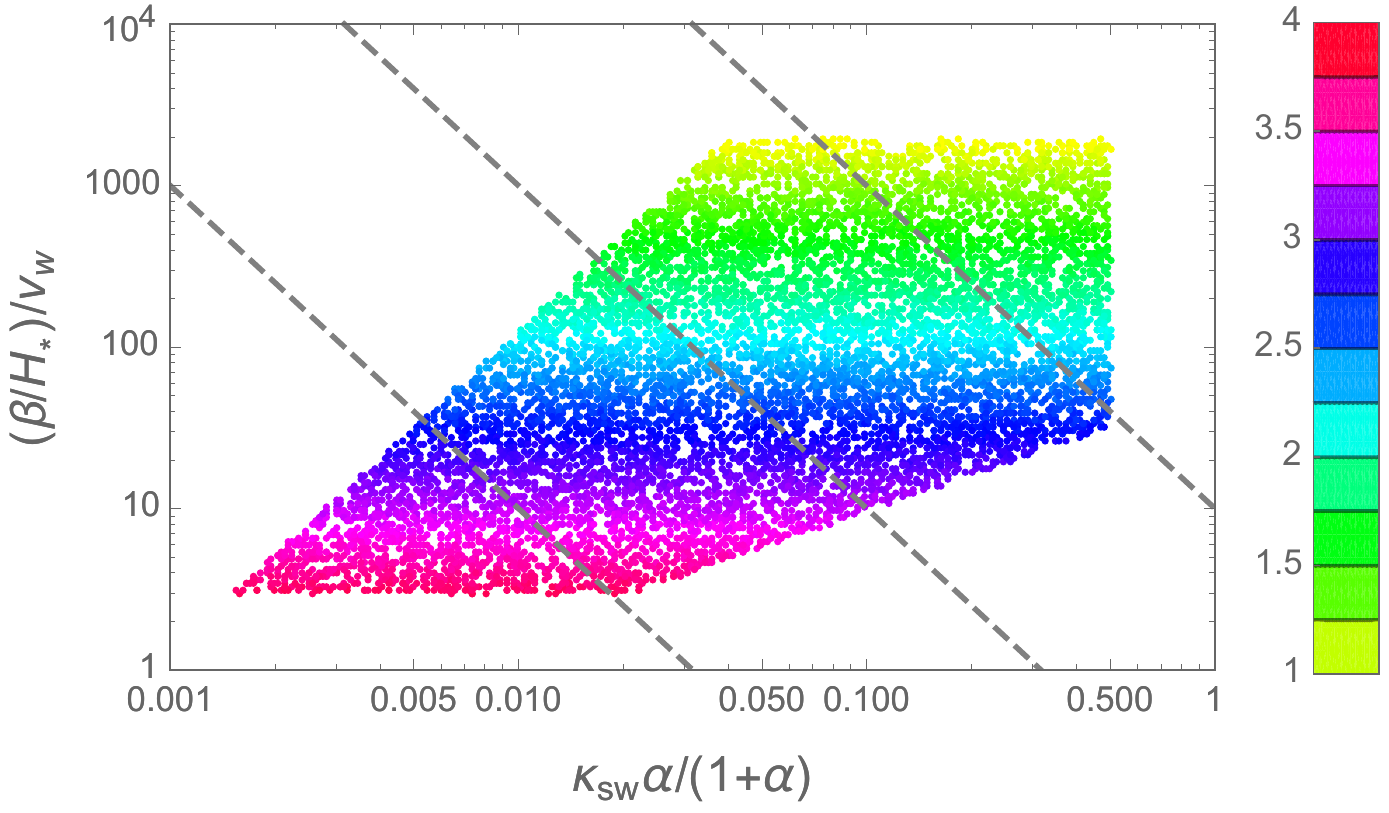}
\includegraphics[width=0.48\textwidth]{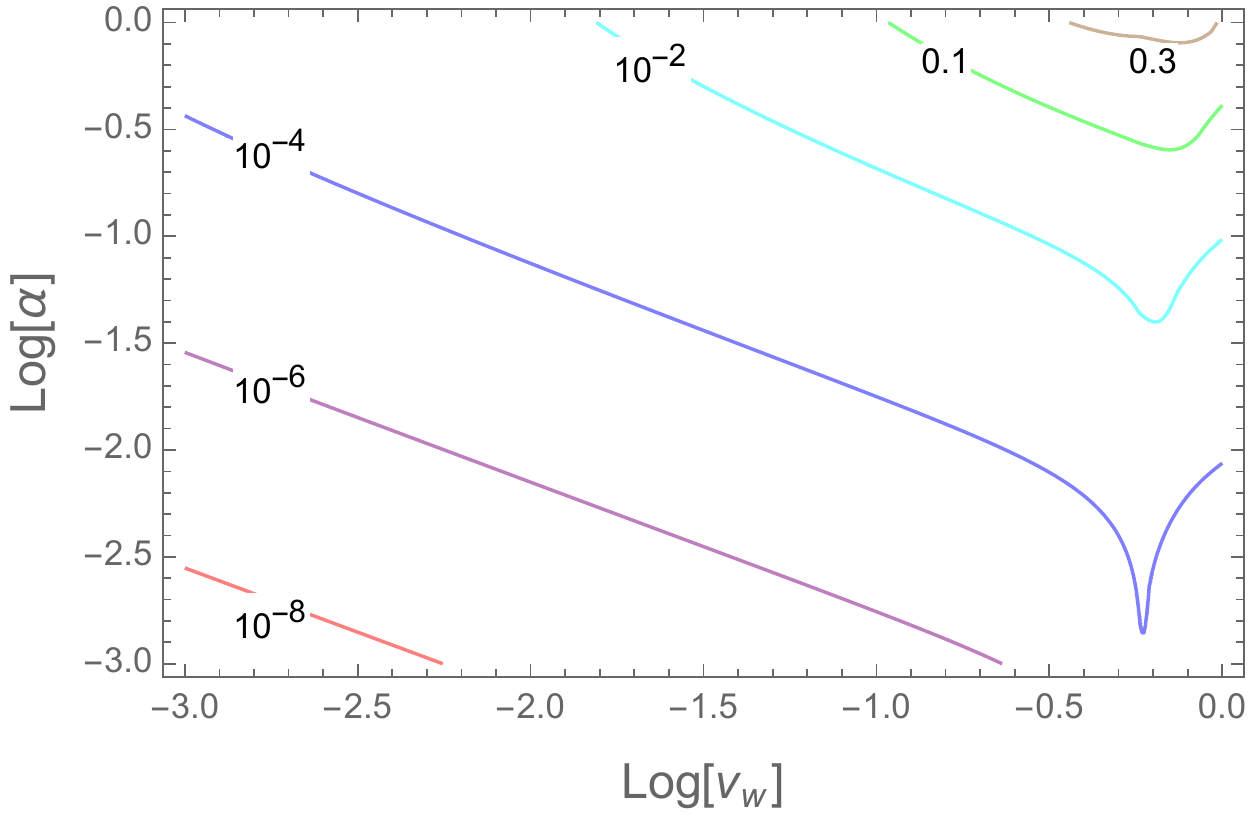}
\caption{Degeneracy in GW spectra in RD for non-runaway bubbles.
\textit{Top}: Parameter space allowing only single contribution (from sound wave) in the reach of LISA.
The labels of color caption indicate $\log(T_*/\GeV)$.
Parameters were scanned, covering $(\beta/H_*)/v_w = \l[3,10^4\r], \ \kappa_{\rm sw} \alpha/(1+\alpha) = \l[ 10^{-4},0.5\r]$ and $T_*/\GeV= \l[10,10^4\r]$.
They were constrained such that $f_{\rm turb}^{\rm peak} = (2.7-3.3) \times 10^{-3} {\rm mHz}$ around the best sensitivity region of LISA and $\Omega_{\rm turb}^{\rm RD} < 2.016 \times 10^{-12} \leq \Omega_{\rm sw}^{\rm RD}$.
The gray dashed diagonal lines are examples of constant $\Omega_{\rm sw}^{\rm RD}$.
\textit{Bottom}: $\kappa_{\rm sw} \alpha/(1+\alpha)$ as a function of $\alpha$ and $v_w$.
}
\label{fig:GW-RD-deg2}
\end{center}
\end{figure}
\begin{figure}[h!]
\begin{center}
\includegraphics[width=0.48\textwidth]{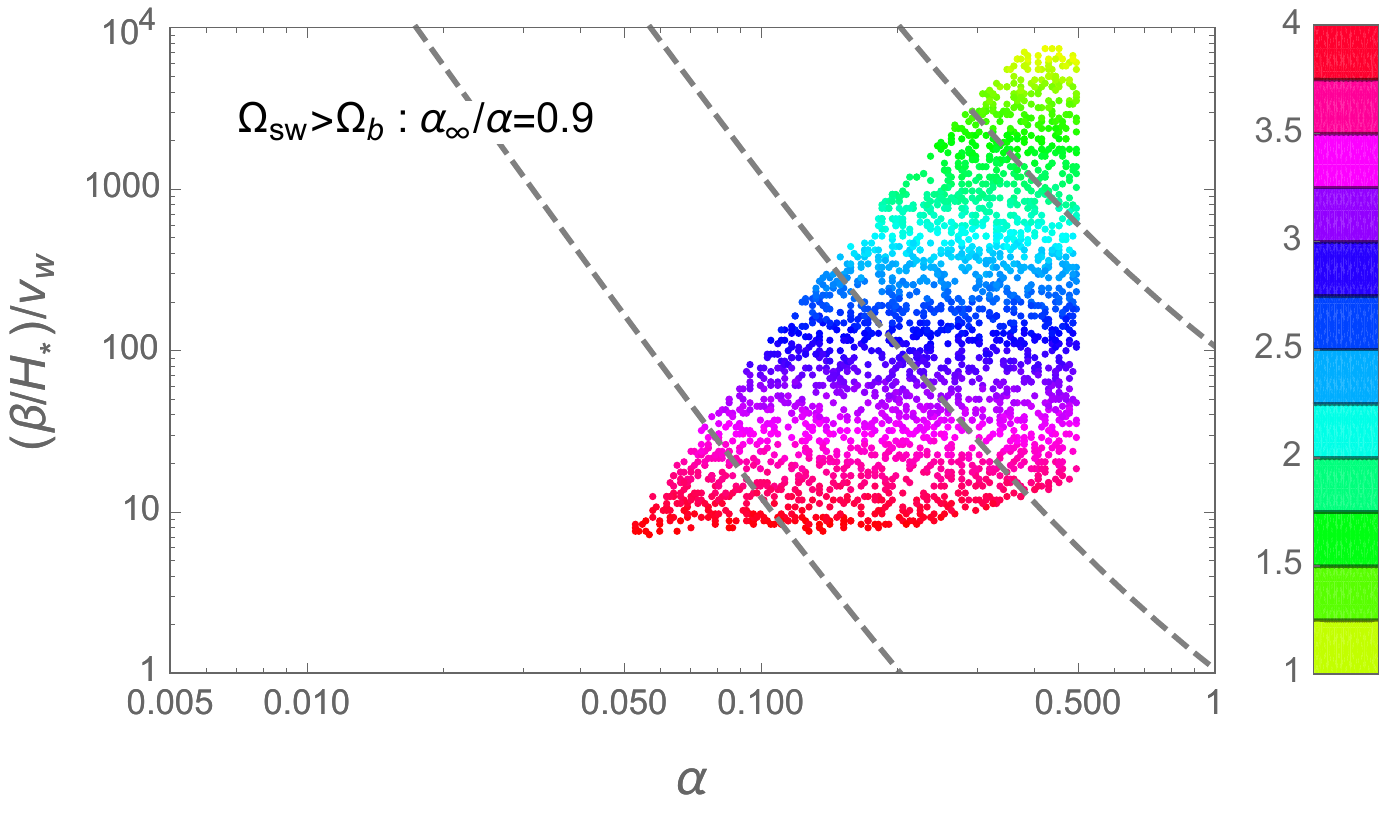}
\includegraphics[width=0.48\textwidth]{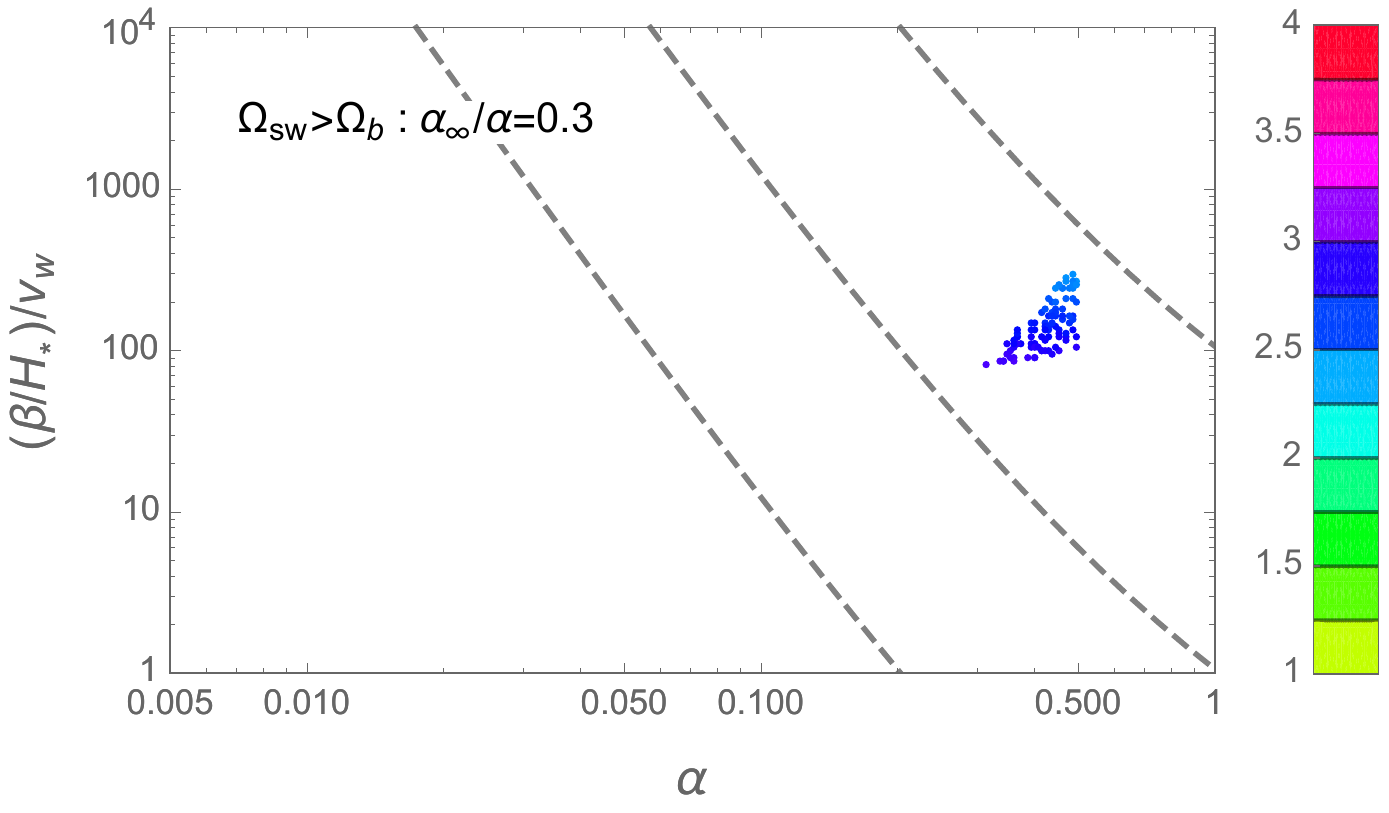}
\includegraphics[width=0.48\textwidth]{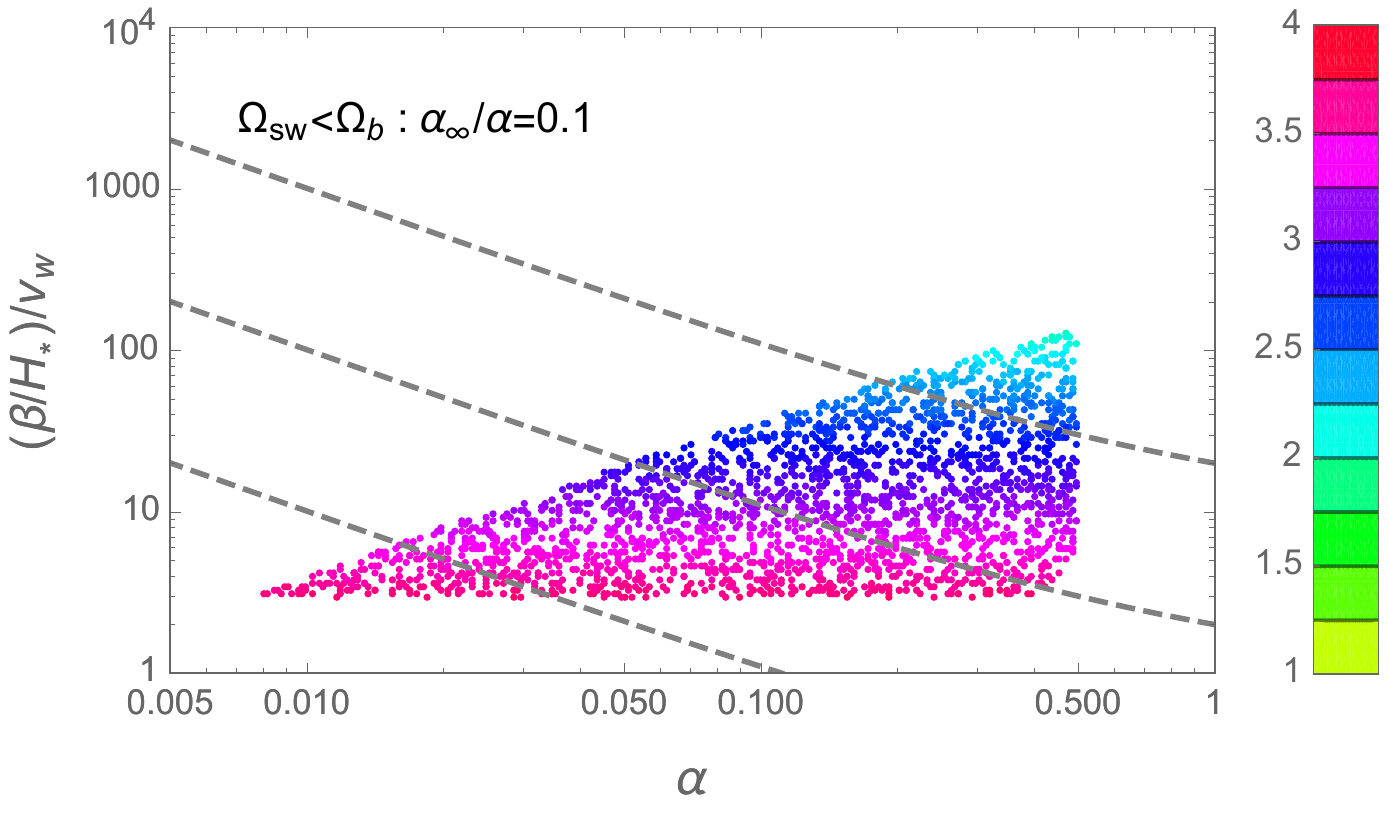}
\caption{Degeneracy in GW spectra in RD for runaway bubbles.
Parameter space allowing only the contribution from bubble collisions or sound waves in the reach of LISA.
Parameters were scanned, covering $(\beta/H_*)/v_w = \l[3,10^4\r], \ \alpha = \l[ 10^{-3},0.5\r]$ and $T_*/\GeV= \l[10,10^4\r]$.
Gray dashed diagonal lines represents contours of the main contribution to the amplitude of GWs. 
\textit{Top}: Parameters were constrained such that $f_{\rm b}^{\rm peak} = (2.7-3.3) \times 10^{-3} {\rm mHz}$ around the best sensitivity region of LISA and $\Omega_{\rm b}^{\rm RD} < 2.016 \times 10^{-12} \leq \Omega_{\rm sw}^{\rm RD}$.
The gray dashed diagonal lines are examples of constant $\Omega_{\rm sw}^{\rm RD}$.
\textit{Middle}: The same as the top panel but with $\alpha_\infty/\alpha=0.3$.
\textit{Bottom}: Parameters were constrained such that $f_{\rm sw}^{\rm peak} = (2.7-3.3) \times 10^{-3} {\rm mHz}$ around the best sensitivity region of LISA and $\Omega_{\rm sw}^{\rm RD} < 2.016 \times 10^{-12} \leq \Omega_{\rm b}^{\rm RD}$.
The gray dashed diagonal lines are examples of constant $\Omega_{\rm b}^{\rm RD}$.
}
\label{fig:deg-RD-ra}
\end{center}
\end{figure}
In particular, although it is upper-bounded, it is difficult to determine the energy scale or $T_*$ at which the GW backgroud is expected to be generated.
Therefore, it is crucial for a detector to have a sensitivity good enough to distinguish subdominant contributions. 

In Fig.~\ref{fig:GW-RD-deg1} and~\ref{fig:GW-RD-deg2}, we show the afore-mentioned degeneracy for non-runaway bubbles in regard of LISA sensitivity \footnote{Depending on its configuration, the sensitivity of eLISA may not be good enough to probe the peak amplitude of the turbulence contribution for the range of macroscopic parameters we considered here as an example.
So, we considered LISA.
}. 
In the upper panel of Fig.~~\ref{fig:GW-RD-deg2}, we used the peak frequency of the turbulence contribution $f_{\rm turb}^{\rm peak}$ which is found to be
\beq
f_{\rm turb}^{\rm peak} \simeq \l( \frac{3 A-1}{5 A} + \frac{9}{3 A-1} \r) f_{\rm turb}
\eeq
where 
\beq
A(\beta/H_*, v_w) = \frac{1.634 \times 8 \pi}{v_w} \l( \frac{\beta}{H_*} \r)
\eeq
and $f_{\rm turb}$ is given by \eq{f-turb}. 
In the panel, one can clearly see that for an expected GW signal macroscopic parameters can span a couple of orders of magnitude, while predicting the same  amplitude of GW signal (for example) at frequencies around the best sensitivity of LISA. 
This degeneracy makes it difficult to pin down the microphysical origin of the GWs.

In the case of runaway bubbles, depending on the ratio $\alpha_\infty/\alpha$, the dominant contribution can either be bubble collisions or sound waves.
As shown in Fig.~\ref{fig:deg-RD-ra} where an analysis similar to the case of non-runaway bubbles was performed, for $\alpha_\infty/\alpha \lesssim 0.3$ bubble collisions are likely to dominate over the contribution from sound waves.
From the top and bottom panels of the figure, we again notice that, although they are weaker than in the case of non-runaway bubbles, there are large degeneracies among parameters when only the main contribution to an expected GW signal is within the reach of a detector.  

Actually, since all the peak frequencies of the relevant contributions are fixed once $(\beta/H_*) T_*/v_w$ is fixed, a determination of all four free-parameters requires at least three contributions within the reach of a detector.
This means that for non-runaway bubbles which have only two relevant sources of GWs (i.e., sound wave and turbulence), it is not possible to determine the macroscopic parameters completely.
However, contrary to runaway bubbles in plasma, in this case it is possible to determine $T_*$ when its two relevant contributions are detected.
This is thanks to a specific correlation of those two contributions:
If $\kappa_{\rm turb}$ is simply proportional to $\kappa_{\rm sw}$ as \eq{kappa-turb} and the proportionality constant is known \cite{Caprini:2015zlo}, one can regard the following set of parameters as free ones.
\beq \label{nra-para}
\frac{\beta}{v_w H_*}, \ \frac{\kappa_{\rm sw} \alpha}{1+\alpha}, \ T_*
\eeq
reducing the number of unknowns to be determined.
Thus, if a detector is sensitive enough to the two contributions, the three pieces of information (peak amplitude and frequency, and a feature in the slope) can completely determine those three parameters.   
Hence, even if the degeneracy among macroscopic parameters in determining the first two parameters in \eq{nra-para} is not lifted, as shown in Fig.~\ref{fig:GW-RD-deg2} for example, it becomes possible to determine $T_*$.
%

In the case of MD, the mode-dependent redshift described in \eq{Omega-GW-0} introduces a unique feature in the spectrum.
Hence, as long as such a feature is detected, even if only one contribution is relevant for a given detector, from \eqs{f-beta}{f-Td} one can find immediately $H_*$ and $H_\times$ (or $\Gamma_{\rm d}$), and the associated $\kappa \alpha/(1+\alpha)$ can be determined.
Still the degeneracy in the combination of $\alpha$ and $v_w$ can not be broken unless at least one more information appears.
So, in the case of matter domination, even if the spectral distortion caused by an additional dilution is detected, it is possible to determine all the macroscopic parameters only for the case of run-away bubbles with at least two contributions within the reach of a detector.
\begin{table}[h]
\begin{center}
\newcolumntype{C}{X<{\centering}}
\begin{tabularx}{0.47\textwidth}{|c||C|C|C||C|C||c|c|}
\hline
& \multicolumn{3}{c||}{Source} & \multicolumn{2}{c||}{\# of free para.} & \multicolumn{2}{c|}{$H_*$} \\
\cline{2-8}
& Bobble & S. W. & Turb. & RD & MD & RD & MD  \\
\hline \hline
\multirow{2}{*}{NRA} & $-$ & $\surd$ & $\times$ & 2 & 1 & $\times$ & $\surd$ \\
\cline{2-8}
 & $-$ & $\surd$ & $\surd$ & 1 & 1& $\surd$ & $\surd$ \\
\hline \hline
\multirow{3}{*}{RA} & $\surd$ & $\times$ & $\times$ & 2 & 1 & $\times$ & $\surd$ \\
\cline{2-8}
 & $\surd$ & $\surd$ & $\times$ & 1 & 0 & $\times$ & $\surd$ \\
\cline{2-8}
 & $\surd$ & $\surd$ & $\surd$ & 0 & 0 & $\surd$ & $\surd$ \\
\hline
\end{tabularx}
\end{center}
\caption{Discrimination power of GWs. ``$\surd$'' and ``$\times$'' denote the possibility of a detector being sensitive to the source or the possibility of determining $H_*$. ``\# of free para.'' denotes the \# of remaining undetermined macroscopic parameters.}
\label{tab:dis-power}
\end{table}%

If the specific feature of MD is not detected, it is difficult to know if a detected signal is generated in MD or RD, since the signal can be obtained in both of MD and RD by a properly chosen set of parameters. 
This adds another degeneracy in a GW signal which may be detected in the future.
Note however that if the duration of phase transition is somewhat long, for example $1 < \beta/H_* \lesssim 100$, the effect of the energy injection from the dominating matter to radiation would be significant or sizable and result in spectral changes.
In this case, a GW signal from a first order phase transition would contain clear information about whether the Universe was dominated by matter or radiation when the signal was generated.
This case will be studied elsewhere.

The discriminating power for both cases of RD and MD is summarized in Table~\ref{tab:dis-power}, assuming the kink-like feature of MD is detected.
In the table, one can see the number of contributions from a first order phase transition that should be detected in order to determine macroscopic parameters governing the peak amplitudes and frequencies.

\section{Conclusions}

In this work, we investigated a stochastic gravitational wave background from a short-lasting first order phase transition in a matter-dominated universe.
Ignoring the effect of energy injection from the dominating matter to radiation, possible interactions between radiation/scalar (of the phase transition) and background matter, we show that the spectrum of the GWB soon after the generation is the same as the one expected in a radiation-dominated universe, and that a mode-dependent additional red-shift during matter-domination era introduces a  unique and distinctive feature which provides important information about the properties of the phase transition and thermal history of the universe.

%
%

We also discussed an inverse problem of a GW signal in view of degeneracies among macroscopic parameters governing the amplitude and spectral shape of the GW signal, showing that wide ranges (covering one or two orders of magnitude) of different sets of parameters can result in a specific GW signal if only the main contribution among the three relevant ones of a first order phase transition is within the reach of detector sensitivity.
For a GW signal generated in an early matter-domination era, if its unique spectral feature is out of detector sensitivities, it is difficult to know whether the signal is generated in a matter-dominated universe or not since the same signal can be generated in a radiation-dominated universe but with a different parameter set.
This adds another degeneracy in a GW signal.
As shown in Table~\ref{tab:dis-power}, a complete breaking of the degeneracy (or determining all the macroscopic parameters) is possible only for the case of runaway bubbles in plasma in both of RD and MD, but when one or two more contributions in addition to the main one are detected.

In regard of $H_*$ the expansion rate around the epoch of the generation of the gravitational waves, in the case of radiation-dominated universe, it can be determined only when all the relevant contributions associated with each bubble dynamics (non-runaway or runaway) can be detected.
This is true for the case of matter-dominated universe if its spectral feature is out of the detector sensitivity.
However, if the spectral feature of matter domination is observed, $H_*$ can be determined always even if there is only a single contribution within the reach of the detector. 

For phase transitions whose durations are not so short relative to $1/H_*$, the energy injection should be taken into account and may cause spectral changes relative to the one from a short-time phase transtion.
This issue will be discussed elsewhere.


\section{Acknowledgements}
The authors are grateful to Geraldine Servant for her helpful comments. 
They also acknowledge support from the MEC and FEDER (EC) Grants SEV-2014-0398 and FPA2014-54459 and the Generalitat Valenciana under grant PROME- TEOII/2013/017. This project has received funding from the European Union's Horizon 2020
research and innovation programme under the Marie Sklodowska-Curie grant
Elusives ITN agreement No 674896  and InvisiblesPlus RISE, agreement No 690575.

\section{Appendix}

The gravitational waves from a first order phase transition are known to consists of three contributions from bubble-collisions, sound-waves, and turbulence such that \cite{Caprini:2015zlo}
\beq
\Omega_{\rm GW} = \Omega_{\rm b} + \Omega_{\rm sw} + \Omega_{\rm turb}
\eeq
where $\Omega_i$ represents the fractional energy:
\begin{widetext}
\bea \label{Omega-b}
h^2 \Omega_{\rm b} &=& 1.67 \times 10^{-5} \l( \frac{H_*}{\beta} \r)^2 \l( \frac{\kappa_{\rm b} \alpha}{1+\alpha} \r)^2 \l( \frac{100}{g_*} \r)^{\frac{1}{3}} \l( \frac{0.11 v_w^3}{0.42+v_w^2} \r) S_{\rm b}(f)
\\ \label{Omega-sw}
h^2 \Omega_{\rm sw} &=& 2.65 \times 10^{-6} \l( \frac{H_*}{\beta} \r) \l( \frac{\kappa_{\rm sw} \alpha}{1+\alpha} \r)^2 \l( \frac{100}{g_*} \r)^{\frac{1}{3}} v_w S_{\rm sw}(f)
\\ \label{Omega-turb}
h^2 \Omega_{\rm turb} &=& 3.35 \times 10^{-4} \l( \frac{H_*}{\beta} \r) \l( \frac{\kappa_{\rm turb} \alpha}{1+\alpha} \r)^{\frac{3}{2}} \l( \frac{100}{g_*} \r)^{\frac{1}{3}} v_w S_{\rm turb}(f)
\eea
\end{widetext}
where 
\bea
S_{\rm b} &=& \frac{3.8 (f/f_{\rm b})^{2.8}}{1+2.8 (f/f_{\rm b})^{3.8}}
\\
S_{\rm sw} &=& \l( \frac{f}{f_{\rm sw}} \r)^3 \l( \frac{7}{4+3 (f/f_{\rm sw})^2} \r)^{\frac{7}{2}}
\\
S_{\rm turb} &=& \frac{(f/f_{\rm turb})^3}{\l[ 1+(f/f_{\rm turb}) \r]^{\frac{11}{3}} \l( 1+8\pi f/h_* \r)}
\eea
with the peak frequency of bubble contribution at the time of GW-production 
\beq \label{fstar-beta}
\frac{f_*}{\beta} = \frac{0.62}{1.8-0.1 v_w + v_w^2}
\eeq
and the inverse Hubble time at GW production, redshifted today in the standard thermal history, 
\beq
h_* = 0.165 \ {\rm mHz} \l( \frac{T_*}{1 \TeV} \r) \l( \frac{g_*}{100} \r)^{\frac{1}{6}}
\eeq
The peak frequencies today are
\bea \label{f-b}
f_{\rm b} &=& 0.165 \ {\rm mHz} \l( \frac{f_*}{\beta} \r) \l( \frac{\beta}{H_*} \r) \l( \frac{T_*}{1 \TeV} \r) \l( \frac{g_*}{100} \r)^{\frac{1}{6}}
\\ \label{f-sw}
f_{\rm sw} &=& 0.19 \ \frac{\rm mHz}{v_w} \l( \frac{\beta}{H_*} \r) \l( \frac{T_*}{1 \TeV} \r) \l( \frac{g_*}{100} \r)^{\frac{1}{6}}
\\ \label{f-turb}
f_{\rm turb} &=& 0.27 \ \frac{\rm mHz}{v_w} \l( \frac{\beta}{H_*} \r) \l( \frac{T_*}{1 \TeV} \r) \l( \frac{g_*}{100} \r)^{\frac{1}{6}}
\eea

The efficiency factors are given as follows \cite{Espinosa:2010hh,Caprini:2015zlo}.

\subsection{Non-runaway bubbles: $\alpha \leq \alpha_\infty$}

In this case, the contribution from bubble collisions is negligible and for the contribution from sound waves
\begin{widetext}
\bea
\kappa_{\rm sw}(v_w \lesssim c_s) &\simeq& \frac{c_s^{11/5} \kappa_A \kappa_B}{\l( c_s^{11/5} - v_w^{11/5} \r) \kappa_B + v_w c_s^{6/5} \kappa_A}
\\
\kappa_{\rm sw}(c_s < v_w < v_J) &\simeq& \kappa_B + \l( v_w-c_s \r) \delta \kappa + \l( \frac{v_w-c_s}{v_J-c_s} \r)^3 \l[ \kappa_C - \kappa_B - \l( v_J - c_s \r) \delta \kappa \r]
\\
\kappa_{\rm sw}(v_J \lesssim v_w) &\simeq& \frac{\l(v_J-1\r)^3 v_J^{5/2} v_w^{-5/2} \kappa_C \kappa_D}{\l[ \l( v_J-1\r)^3 - \l(v_w-1\r)^3 \r] v_J^{5/2} \kappa_C + \l( v_w-1 \r)^3 \kappa_D}
\eea
\end{widetext}
where 
\bea
\kappa_A &\simeq& v_w^{6/5} \frac{6.9 \alpha}{1.36 - 0.037 \sqrt{\alpha} + \alpha}
\\
\kappa_B &\simeq& \frac{\alpha^{2/5}}{0.017+ \l( 0.997 + \alpha \r)^{2/5}}
\\
\kappa_C &\simeq& \frac{\sqrt{\alpha}}{0.135+\sqrt{0.98+\alpha}}
\\
\kappa_D &\simeq& \frac{\alpha}{0.73+ 0.083 \sqrt{\alpha}+\alpha}
\\
\delta \kappa &\simeq& - 0.9 \log \frac{\sqrt{\alpha}}{1+\sqrt{\alpha}} 
\eea
and the Jouguet velocity is 
\beq
v_J = \frac{\sqrt{2 \alpha/3 + \alpha^2}+\sqrt{1/3}}{1+\alpha}
\eeq
Although it would have to determined by an appropriate numerical simulation, one may set \cite{Caprini:2015zlo} 
\beq \label{kappa-turb}
\kappa_{\rm turb} = \epsilon \kappa_{\rm sw}
\eeq
with $\epsilon = \mathcal{O}(0.1)$.

\subsection{Runaway bubbles in a plasma: $\alpha > \alpha_\infty$}

In this case, all the three contributions can be relevant and 
\bea
\kappa_{\rm b} &=& 1- \frac{\alpha_\infty}{\alpha} \geq 0
\\
\kappa_{\rm sw} &=& \frac{\alpha_\infty}{\alpha} \kappa_\infty
\\
\kappa_{\rm therm} &=& \l( 1 - \kappa_\infty \r) \frac{\alpha_\infty}{\alpha} 
\eea
where
\bea
\alpha_\infty &\simeq& \frac{10}{8 \pi^2} \frac{\sum_i c_i \Delta m_i^2(\phi_*)}{g_* T_*^2}
\\
\kappa_\infty &\equiv& \frac{\alpha_\infty}{0.73+0.083 \sqrt{\alpha_\infty}+\alpha_\infty}
\eea
For turbulence, \eq{kappa-turb} is expected to be applicable.

%

\end{document}